\documentclass{article}
\usepackage{arxiv}
\usepackage{geometry}
\usepackage{hyperref}
\usepackage{cite}

\usepackage{graphicx}
\graphicspath{{figures/}{pictures/}{images/}{./}} 

\usepackage{amsmath}
\usepackage[ruled,vlined]{algorithm2e}
\usepackage{algpseudocode} 
\usepackage{array}
\usepackage{microtype}
\usepackage{caption}
\usepackage{booktabs}           

\newcommand{\etal}{\emph{et al.}~}

\title{SynopSet: Multiscale Visual Abstraction Set for Explanatory Analysis of DNA Nanotechnology Simulations}

\usepackage{authblk}
\author[1]{Deng~Luo}
\author[1]{Alexandre~Kouyoumdjian}
\author[1]{Ond\v{r}ej~Strnad}
\author[2]{Haichao~Miao}
\author[3]{Ivan~Barišić}
\author[1]{Ivan~Viola}

\affil[1]{\textit{\small King Abdullah University of Science and Technology (KAUST), Saudi Arabia. E-mails: \{deng.luo $\vert$ alexandre.kouyoumdjian $\vert$ ondrej.strnad $\vert$ ivan.viola~\}@kaust.edu.sa. D. Luo and A. Kouyoumdjian are co-first authors.}}

\affil[2]{\textit{Center for Applied Scientific Computing, Lawrence Livermore National Laboratory, United States. E-mail:  miao1@llnl.gov.}}
\affil[3]{\textit{Center for Health and Bioresources, Austrian Institute of Technology, Austria. E-mail: ivan.barisic@ait.ac.at.}}

\date{}                     
\begin{document}
\maketitle

\begin{abstract}
We propose a new abstraction set (SynopSet) that has a continuum of visual representations for the explanatory analysis of molecular dynamics simulations (MDS) in the DNA nanotechnology domain. By re-purposing the commonly used progress bar and designing novel visuals, as well as transforming the data from the domain format to a format that better fits the newly designed visuals, we compose this new set of representations. This set is also designed to be capable of showing all spatial and temporal details, and all structural complexity, or abstracting these to various degrees, enabling both the slow playback of the simulation for detailed examinations or very fast playback for an overview that helps to efficiently identify events of interest, as well as several intermediate levels between these two extremes. For any pair of successive representations, we demonstrate smooth, continuous transitions, enabling users to keep track of relevant information from one representation to the next. By providing multiple representations suited to different temporal resolutions and connected by smooth transitions, we enable time-efficient simulation analysis, giving users the opportunity to examine and present important phases in great detail, or leverage abstract representations to go over uneventful phases much faster. Domain experts can thus gain actionable insight about their simulations and communicate it in a much shorter time. Further, the novel representations are more intuitive and also enable researchers unfamiliar with MDS analysis graphs to better understand the simulation results. We assessed the effectiveness of SynopSet on 12 DNA nanostructure simulations together with a domain expert. We have also shown that our set of representations can be systematically located in a visualization space, dubbed SynopSpace, composed of three axes: granularity, visual idiom, and information layout type. Proposed at the end are general guidelines on the use of SynopSpace to generate a new SynopSet for the visualization of other dynamic processes involving hierarchical structures.
\end{abstract}


\begin{figure*}
  \centering
  \includegraphics[width=1.0\linewidth]{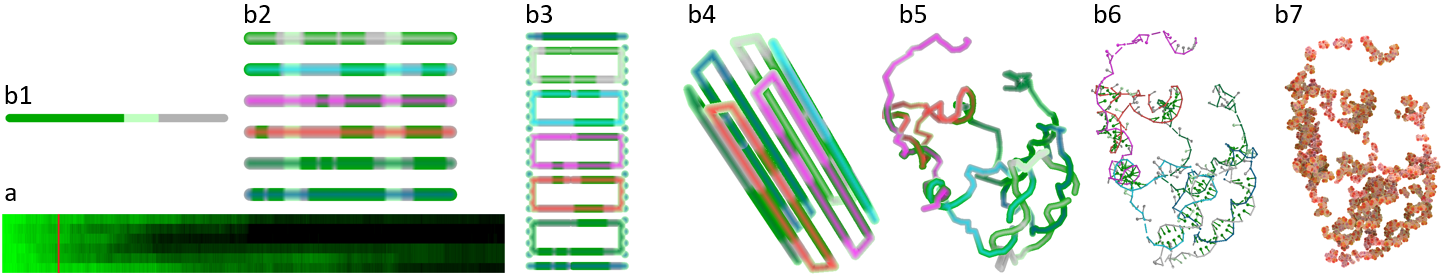}
  \caption{SynopSet, novel visual abstractions for inspecting molecular dynamics simulations (MDS): (a) \textbf{SynopBond figure} is a summary view characterizing the bonding status of each strand over time. (b1-b7)  \textbf{SynoPoints} visualizations are representing a single simulation frame, depicted by the red vertical line in the summary view, in the disassembly of a nanopore structure composed of DNA strands. These \textbf{SynoPoints}, exhibiting different abstraction levels, facilitate the understanding of the hierarchical structure and of the dynamic behavior of MDS. They are sorted from most abstract (left) to most detailed (right) visual representations. All representations show that one side of the pore gets disassembled, due to the breaking of the hydrogen bonds on the red and purple strands with their complementary strands.} 
  \label{fig:teaser}
\end{figure*}

\section{Introduction}\label{sec:introduction}

DNA nanotechnology~\cite{seeman1982nucleic} uses DNA in its double helix form as building blocks for more complex structures, by leveraging the Watson-Crick base pairing rule, rather than as a genetic information carrier. Two main methods have been developed to assemble DNA nanostructures: scaffolded DNA origami~\cite{Rothemund:2006:dnaorigami} and DNA bricks~\cite{wang2017practical, seeman2017dna, ke2012three}. The assembled structures are ever-increasing and range from simple geometric primitives to large multi-component systems with dynamic behavior. Well-designed nanostructures have shown promising application scenarios, ranging from fundamental studies by acting as substrates for biochemical analysis~\cite{rajendran2012single}, to medical applications by facilitating targeted drug delivery~\cite{mela2020dna}. 


A typical current workflow in this domain is to first design the shape of the nanostructure with computer-aided design tools. These will usually be able to generate specific DNA sequences for all the strands in the designed structure. Then the structure will be subject to molecular dynamics simulation to investigate its structural stability and dynamic behavior; if it is not stable, or lacks desired properties, several rounds of modifications followed by simulation loops will be performed until the designer is satisfied. Then the finalized DNA sequences will be ordered as oligonucleotides from commercial companies and lab experiments will be performed to assemble the designed structure. Finally, tests will be run, such as observing the shape of the final assembled structure under the microscope.

DNA simulations may include tens of thousands of atoms, and many thousands of time steps. If a domain expert wishes to analyze a simulation in atomistic detail through visualization, they will therefore be presented with a great deal of information per step. Visualizing all time steps in a continuous playback presents a vastly greater challenge still, for the sheer number of time steps means that in total, there are thousands of positions to analyze over millions of simulation steps, resulting in an enormous amount of data. Fast playback can in principle keep the analysis reasonably short, but in practice, this means a very large amount of information per second will be presented to the expert, vastly exceeding their ability to comprehend it and make use of it. Playing the simulation at a sufficiently low speed to understand it entails many hours of playback, sometimes more. Yet MDS often exhibit long periods devoid of any important events, which would make such a visualization process not only long and tedious, but very inefficient. Such simulations are sometimes visualized at a slightly coarser granularity, by representing only nucleotides and not individual atoms, which somewhat reduces the amount of information per frame, but falls far short of fully solving the problem.

A suitable solution to this problem would provide experts with abstract visualizations to efficiently analyze uneventful phases of their simulations, while still getting an overview of the simulation's state and its dynamics; but it also needs to provide them with very detailed visualizations for critical phases of the simulation. It would also need to provide smooth, continuous transitions between these representations, for a seamless and time-efficient experience.

To solve this, we take advantage of the hierarchical nature of DNA structures to propose involves seven different representations, ranging from very abstract to very detailed. The more abstract ones present less information per simulation step, and involve less (or no) motion of the visual elements themselves, aside from their colors. This makes such representations comfortable to view at high playback speeds, and thus suited to the relatively uneventful phases of DNA simulations. They provide an overview of what happens over a long period of simulation time, but in a short amount of visualization time. On the other hand, they present insufficient detail for the analysis of important events, where experts need detailed information to gain insight about their DNA systems. For such cases, more detailed representations are required, which in turn require lower playback speeds. In summary, the contributions of our solution are:

\begin{itemize}
  \item A new abstraction set (\textbf{SynopSet}) that has a continuum of visual representations for the explanatory analysis of MDS for DNA nanotechnology.
 \item For each representation, i.e., each \textbf{SynoPoint} in \textbf{SynopSet}, we lay out the design rationale and the use of the \textbf{Ballchain} technique used in 5 different SynoPoints. We also demonstrate smooth and continuous transitions between consecutive SynoPoints. 
 \item We show the effectiveness of this novel approach to visual organization and information abstraction 
 on the analysis of seven DNA nanotechnology simulations, generate a summary video, and report the feedback from an expert in DNA nanotechnology.
 \item Finally, we observe that all \textbf{SynoPoints} in \textbf{SynopSet} can be located in a visualization space dubbed \textbf{SynopSpace}, composed of three axes: granularity, visual idiom, and information layout type. We also propose general guidelines on how \textbf{SynopSpace} can be used to generate a new \textbf{SynopSet} for the visualization of other dynamic processes involving hierarchical structures.
\end{itemize}

\section{Background}

At the core of each DNA nanostructure are the hydrogen bonds between Watson-Crick base pairs~\cite{fonseca2000hydrogen}. Among four nucleotide types (commonly abbreviated A, T, C, G), the Watson-Crick rule specifies that A binds to T and C binds to G to form base pairs, stabilizing the nucleic acid strands. Designers can choose the length and sequence of the nucleic acid strands. The DNA origami and DNA bricks methods rely on these properties. DNA origami uses multiple short staple strands 
to fold a very long scaffold strand 
by having consecutive parts of the staple bind to nonconsecutive parts of the scaffold, forming the designed shape of the final structure. DNA bricks, on the other hand, only uses short strands, called bricks, to form the final structure through careful design of the binding pattern between bricks. Thus, both methods involve interactions on multiple scales: hydrogen bonding happens at the atomistic scale, causing the bonding of each base pair (a nucleotide-scale phenomenon), causing the zipping of nucleic acid strands to form a double helix, finally causing the formation of the designed and usually complex final structure. This hierarchical multiscale property will be manifested throughout the design, simulation, and visualization of DNA nanostructure and very often poses technical challenges.

The dynamics of DNA nanostructures involve phenomena that unfold at very small spatial and temporal scales, such as when a hydrogen bond forms between two individual atoms, over a distance of about 0.3nm, in approximately 30ps~\cite{Sheu12683}. But this also involves phenomena taking place at much larger spatial and temporal scales, for instance when a pair of DNA strands bond together or separate from each other, which happens over the entire length of the strand, up to tens of nm and beyond. Therefore, both atomic-level~\cite{maiti2006atomic} and polymer-level~\cite{bustamante1994entropic} simulations can be useful. A very popular simulator in DNA nanotechnology, oxDNA~\cite{snodin2015introducing, vsulc2012sequence, rovigatti2015comparison, Ouldridge:2011:SMT}, runs at the nucleotide level. It is thus capable of capturing the movement of each nucleotide, the binding of Watson-Crick base pairs. It can also capture zipping events between complementary strands because the coarse level allows for sufficiently long simulated times---microseconds or more~\cite{snodin2016direct}. We therefore picked oxDNA as the source of our simulation data sets.

\section{Related work}

\subsection{Molecular Dynamics Visualization}
To visualize the dynamic behavior of molecules, it is important to represent the trajectories resulting from molecular dynamics simulation. Byska \emph{et al.}~\cite{byvska2015animoaminominer} visualize protein tunnels by representing the path of each amino acid in the tunnel and aggregating the trajectories into profiles in this way. Koles\'ar \emph{et al.}. propose a three-level system for illustrating the process of polymerization~\cite{kolesar2014interactively} coupling together, an L-system with agent-based simulation, and quantitative simulation techniques. In further work 
Koles\'ar et al.~\cite{kolesar2016fractional} propose a way to rectify the simulated data to allow for comparative visualization of a cohort. A more general approach to particle-based spatiotemporal data visualization was proposed by Palenik \emph{et al.}~\cite{palenik2019scale}, which enables rapid identification of patterns by simultaneously exploring temporal and spatial scales.

\subsection{DNA Nanotechnology Modeling and Visualization}
Several computer-aided design tools have been developed for DNA nanotechnology, such as Adenita~\cite{de2020adenita}, cadnano~\cite{douglas2009rapid}, and oxView~\cite{poppleton2020design}.
Cadnano uses a lattice of parallel nucleic acid helix strands as base plate and then lets users select active strands and edit the connections between them. The tool's visual interface has all the editing in 2D. This allows easy interactions, but hinders the mental visualization and understanding of the structure in 3D.
The oxView tools offers design functionalities in a web browser. Its interface also allows the visualization of a dynamic simulation. Its ability to render many nucleotides in 3D in a browser is an important advantage, but the lack of abstract views for different scales is a weakness. This makes it difficult for users identify and understand relevant simulation events buried in vast amounts of data, with erratic animation behavior and visual occlusion.

\subsection{Event-adaptive Visualizations}
For a meaningful event or a series of them to happen in MDS, much longer simulation times are required than for MDS used to test the stability of structures. In such cases, only few events are critically important to the observer, and they are sparsely dotted throughout the whole simulation. This presents a visualization challenge: on how do we quickly locate and focus on the relevant data for those events? For the analysis of long protein MDS, Byska \emph{et al.}~\cite{Byska:2019:ALM} adaptively change the spatiotemporal context depending on the importance of the events. Woodring and Shen~\cite{Woodring:2009:MTA} provide a more global approach to time series data using wavelet transforms to produce multiple temporal resolutions. Patro~\cite{patro2010saliency} measures salience and keyframe changes per atom to extract summary overviews that highlight important events in MDS. The MolPathFinder tool~\cite{Alharbi:2016:MIM} provides users with the ability to filter paths of atoms and highlight key events through static representations. While these approaches find ways to highlight important events, they do not adjust the representation to the different phases of MDS. 

\subsection{Visualization Spaces}

All the SynoPoints in our \textbf{SynopSet} can be located in a formal visualization space. Previous research has shown the usefulness of such organization, which Viola and Isenberg call abstraction spaces~\cite{viola2017pondering}. For molecular visualizations, Zwan \emph{et al.}~\cite{Zwan:2011:IMV} and Lueks \emph{et al.}~\cite{Lueks:2011:SCC} propose a multidimensional dimensional space that organizes the visualization along axes such as structure, illustrativeness, and spatial perception. In contrast, Mohammed \etal\cite{Mohammed:2018:AVT} and Miao \etal\cite{Miao:2018:DimSUM} use their abstraction spaces as interactive panels that allow users to transition smoothly from one representation to another. While the former assigns an axis for each structure of interest, the latter proposes a more general approach that organizes representations along aspects of interest, such as layout and scale. The efficacy of animated transitions between different representations have been studied by Heer and Robertson~\cite{Heer:2007:ATS}. While these works have taken initial steps into general visualization spaces, they offer no insight into the incorporation of animation over time, which is required in our case.

\section{Visual Abstractions for Simulations}
\label{Representations in SynopSet}

Populating a suitable set of representations spanning the spectrum from most abstract to most detailed presents a number of challenges. The most detailed representation needs to communicate all the information a standard molecular dynamics simulation may produce: the position of each atom at any given time. The most abstract may only convey the level of completion of the assembly process. For this simple purpose, in similar situations, a progress bar is commonly used. How should the gap between these two extremes be bridged? The requirements are for multiple representations providing different levels of abstraction, but also exhibiting enough similarity between them to allow for continuity during transitions, lest users lose track of the information they are analyzing. Miao \emph{et al.} proposed several such representations~\cite{miao2017multiscale}. However, these were designed for modeling purposes, and as such are only suitable for static models, not dynamic ones, which simulations require. Further, they do not address the need to convey information about the status of the simulation at a given time. Due to their focus on modeling, they provide no option to display an overview of the simulation's progress while abstracting most of the geometric detail. Efficiency is also an important constraint of ours, given that some DNA simulations can involve many thousands of nucleotides, and require smooth animation for comfortable analysis.
Given these requirements, we extended the versatile technique introduced by Mindek \emph{et al.} to render mitochondria~\cite{mindek2017visualization}. We use impostor spheres to create an elongated, curvy shape akin to flexible tubes, as in figure~\autoref{fig:teaser}b4. We generate it from a set of control points along which we build a spline. We then place impostor spheres on this spline, with sufficient density to create the appearance of a continuous solid. Then, we apply a trilateral filter on the normal buffer, and perform the final shading and ambient-occlusion in screen-space. We call this technique Ballchain.

In most representations, to convey more information, we rely on a dual coloring approach, where each section of this shape features one color on its inner side, and another one on its contour. We achieve this with a post-processing filter 
that generates a halo effect around the Ballchains.

This technique is used for five of the representations presented in this paper, in ways that differ for each representation, as detailed further below. It is particularly useful given the hierarchical nature of DNA assembly. Indeed, an entire structure is made up of several DNA strands. Each DNA strand may contribute to several (typically straight) structural elements. Each portion of strand in a given structural element is made up of nucleotides, and each nucleotide contains several atoms. Our approach therefore relies on representations that focus on distinct hierarchical levels.

\begin{figure*}
  \centering
  \includegraphics[width=0.9\linewidth]{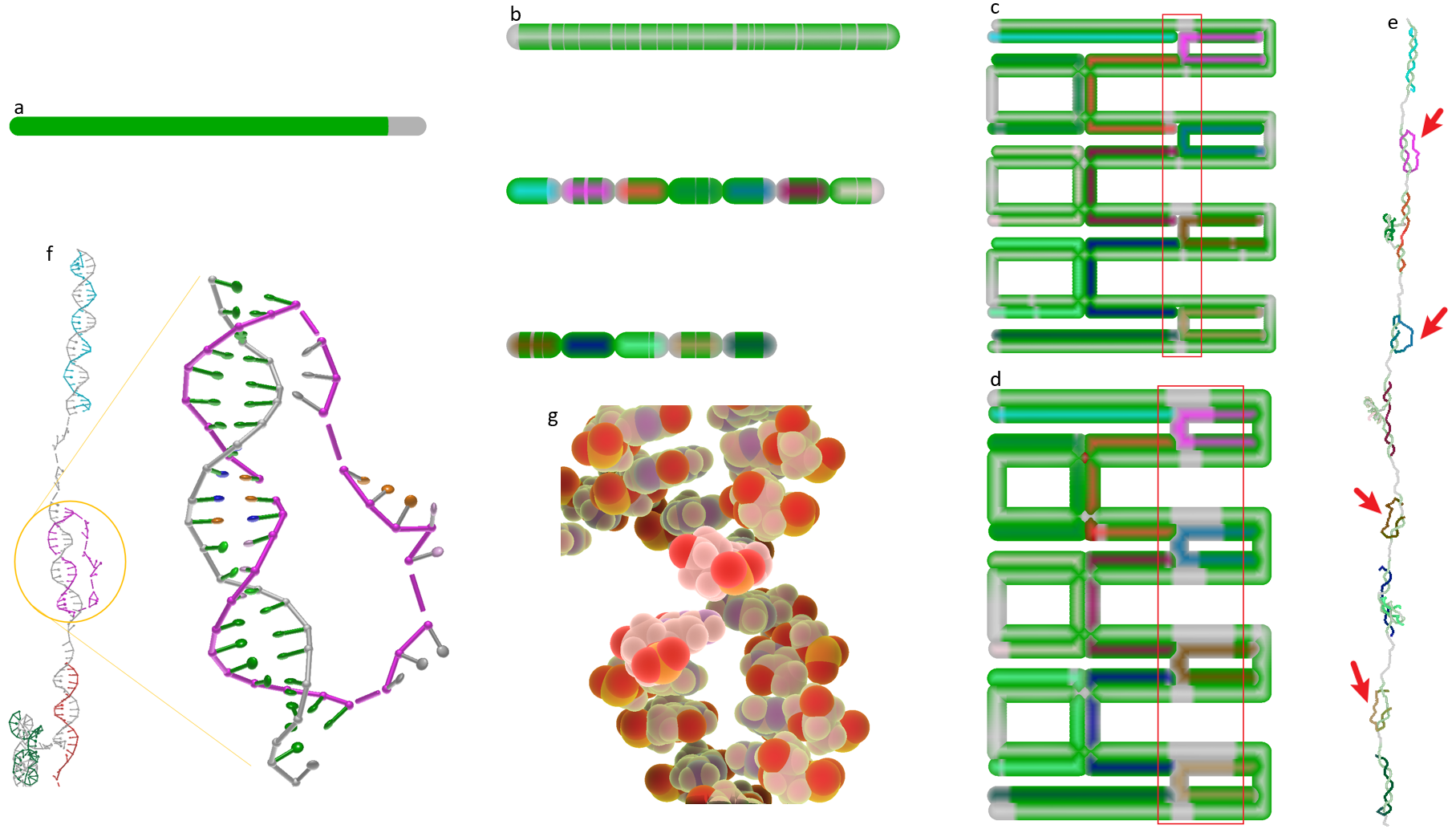}
  \caption{\textbf{SynopSet} representations for \textbf{Snodin's Plane} in \autoref{Results}). (a) Progress Bar; (b) Heatbar Array; (c) Schematic-2D; (d) Schematic-3D; (e) Precise Snakes; (f) All-NT; (g) Atomistic} 
	\label{fig:SynopSet}
\end{figure*}

\textbf{Progress Bar} provides an overview of the assembly process for the entire structure, showing its completion status at any given time, with respect to the final target design. We compute the proportion of correctly paired bases and display it in green, of incorrectly paired bases and display it in red, and of unpaired bases, displayed in gray. For nucleotides that are not bonded, but are not meant to be bonded in the final design, we use a lighter shade of green. The bar is colored sequentially from left to right, with the green portion and light green portion on the left, the red portion on the right, and the gray portion in the middle (see~\autoref{fig:teaser}b1, \autoref{fig:SynopSet}a). 
In keeping with common practice for progress bars, we use flat shading to give the bar the appearance of a 2D object.
This representation is the most abstract one, providing users with basic information about the assembly's status at a glance. Since the bar itself does not move, though its colors can change, it can be used comfortably at very high playback speeds.
  
\textbf{Heatbar Array} displays the status of each individual strand. Here, each strand is represented by a Heatbar (a 1D heatmap), where the inside of the bar is colored so as to uniquely identify the corresponding strand, while the contour is colored to represent the bonding status of each nucleotide, using the same coloring scheme as in the Progress Bar. The strands are presented in a 2D array on the screen, in the order provided by the user. As a heatbar's colors show the status of nucleotides in their actual order, using successive stripes, this representation provides bonding information about the pairing status of each nucleotide, for each strand, in a compact, sequential format. The Progress Bar, in contrast, aggregates all of this information into continuous single color blocks. Because here, each of the heatbars stands for an actual strand, we use 3D shading, as a hint that these represent physical (if nanoscopic) objects. Examples are shown in~\autoref{fig:teaser}b2, \autoref{fig:SynopSet}b. There is more information shown on the screen than with the progress bar, with more significant changes in the colors as the animation is playing, but the layout remains simple, and there is no motion of the geometry. Therefore, this representation can be used at high speeds comfortably, albeit slower than the Progress Bar.
  
\textbf{Schematic-2D} presents a 2D layout of the structure, a planar overview that hints at its 3D shape, much like the net of a polyhedron. This representation is still made up of heatbars identical to those of the Heatbar Array, but with a layout that better represents the structure's shape. In particular, the various bends featured on a strand are represented, such that the heatbar that represents each strand may feature a number of angles. The main benefit of this representation is that although structural information is conveyed, the status of every single nucleotide remains visible in the same view, since the representation is purely 2D, "exploded", and therefore free of any occlusion.
The layout is provided by the user as a text file containing, for each strand, its beginning, its end, and any point where it may bend (with the corresponding nucleotide-ID). DNA nanotechnology design tools may, in some cases, provide this information. We use inverse projection to compute the required 3D coordinates for the impostor spheres; once re-projected in the graphics pipeline, they will reproduce the intended 2D layout. This is necessary due to our use of a perspective camera, which is itself useful to allow for smooth transitions into the following SynoPoint, which is 3D in nature.
Examples are shown in \autoref{fig:SynopSet}c (DNA origami), and \autoref{fig:teaser}b3 (DNA bricks). Since the layout is more complex than that of the Heatbar Array, with a greater number of straight segments, the amount of visual information is higher. However, the representation remains static, with only color changes, so high playback speeds remain comfortable, though to a lesser extent than with the Heatbar Array.
  
\textbf{Schematic-3D} displays the idealized target geometry, into which the DNA strands are supposed to assemble. Each strand remains represented by a heatbar made up of straight segments with sharp bends, but in 3D, with a layout that closely resembles the structure's target shape. Examples are shown in~\autoref{fig:teaser}b4 and~\autoref{fig:SynopSet}d. The advantage over the Schematic-2D representation is that the shape is better conveyed, making it easier for users to get a sense of where each strand fits in the final structure. On the other hand, and especially with complex structures, some strands may be occluded, meaning that users may need to rotate the structure in order to observe what they wish to. Moving the camera closer or even inside the structure may also be required, for instance if inner parts of it are occluded from all angles. Due to this potential need to rotate the structure and manage the camera, comfortable playback speeds will be kept lower than in Schematic 2D; but as the structure itself is not animated, this speed can remain somewhat intermediate.
  
\textbf{Precise Snakes} show the strands in their locations in time, conveying their actual shape, as provided in the simulation's output. It is the first representation with an animated structure, as opposed to just animated colors. Here, the strands are represented with flexible solids of circular section (``Snakes''), which shift and bend in time to match the positions of the nucleotides that make up the strands. This representation offers users who have more time on their hands the opportunity to examine the simulation in more spatial detail. Since the position and shape of a given strand at a given time are determined by the precise positions of its constituent nucleotides, if two strands are paired, their proximity and respective shapes will result in the typical double helix representation, as in~\autoref{fig:teaser}b5 and \autoref{fig:SynopSet}e. Further, the bonding status of the strands is still shown in the contours of the Snakes, while the colors used for their inner parts still serve as unique identifiers for the strands.
As the first representation with spatial animation, Precise Snakes requires significantly lower playback speeds to be viewed comfortably, but as the most abstract animated representation, featuring no structural detail beyond the shape of each strand, it is the animated representation that is best suited to relatively high playback speeds. To make such speeds more comfortable, we apply temporal smoothing to the simulation's output for this representation, performing a sliding window averaging of each nucleotide's position over time, as detailed in section~\autoref{sec:smoothing}. Feedback from domain experts helped us calibrate the width of the sliding window, to reach a good compromise between smoothness and the preservation of the overall dynamics of the simulation, allowing for much higher playback speeds than would otherwise be comfortable.
  
  
\textbf{All-NT} matches the granularity of oxDNA's simulations, conveying all the information in the simulation's output. It communicates the precise position and orientation of each nucleotide in time. We use sticks to represent a DNA strand's backbone with spheres marking the phosphate groups, and ellipsoids to depict the bases, which is a common way~\cite{couch2006nucleic} to ensure that base orientation is communicated, not just their position. These ellipsoids are also connected to the backbone by sticks. All of these shapes are implemented with triangle meshes rendered with instancing. Examples are shown in~\autoref{fig:teaser}b6 and \autoref{fig:SynopSet}f. We use the same overall coloring scheme as with the Snakes, except that the colors used for strand identification are on the backbone of each strand, whereas the bonding status is indicated by coloring the nucleotides themselves (and the sticks that connect them to the backbone), since this is where bonding actually happens.
Beyond communicating the position and orientation of the nucleotides, this representation can indicate the type of each nucleotide, through colors: when the camera gets close to a given nucleotide, its color changes from the green/gray/red indicating bonding status to one of 4 colors indicating its nature (A, T, C, or G). The stick connecting it to the backbone remains colored according to the bonding status, preserving this information.
With the position, orientation, or even nature of each nucleotide shown, the amount of information conveyed in All-NT is significantly higher than with Precise Snakes. As such, it is particularly useful at lower playback speeds, which is why we also chose to disable temporal smoothing, providing a very precise depiction of the conformation of the structure in time.
  
\textbf{Atomistic} displays the Van der Waals surfaces~\cite{bondi1964van} of all atoms in their exact positions in time, without any temporal smoothing. This is a very common SynoPoint for atomistic structures, where each atom is depicted by a sphere, and it is our most detailed representation, implemented with simple impostor spheres. In practice, current DNA simulation tools such as oxDNA are often limited to nucleotide granularity, so we simply deduce the positions of the atoms from the position and orientation of the nucleotides to which they belong, whose structures are known. The motion of all atoms within a nucleotide is therefore fixed in our case, but this would not be so with output from atomistic simulations, where all atoms could move relative to one another, even within a given nucleotide. This SynoPoint is the most appropriate one to envision potential chemical optimization of the structure, in order to facilitate a successful assembly. Examples are shown in~\autoref{fig:teaser}b7 and \autoref{fig:SynopSet}g. Each atom is colored with a blend of the color used to identify its strand, and a color associated with its nature---carbon, hydrogen, oxygen, etc. In addition, each atom features a ``halo'' whose color indicates the bonding status of its associated nucleotide. 
Given the very high amount of information displayed, this SynoPoint is best suited to very low playback speeds, for users looking to examine things in very fine detail.

A summary figure called \textbf{SynopBondFigure} is designed and shown in \autoref{fig:teaser}a to further facilitate the understanding of the whole dynamic process. It is inspired by and adapted from Figure 3 from Engal \emph{et al.}'s work~\cite{engel2018force}. Each horizontal lane represents one strand, and each vertical column represents one step in the MDS. The coloring encodes the number of bonded nucleotides on that strand, where deeper green indicates more bonds. The vertical red line, which users can move, indicates the current step shown in the \textbf{SynopSet} SynoPoints. Although the design of this visual encoding is not novel, the coupling of such a summary figure with all the main per-step representations (SynoPoints) is, and enables the quick identification of interesting events in long simulations. There are other figures in this type that can be coupled too. We propose the term \textbf{SynopFigure} to encompass all figures that represent time-unrolled views. Our SynopBondFigure is therefore but one SynopFigure out of many possible ones.

\section{Smooth Transitions Between SynoPoints}

When domain experts wish to get a quick overview of the simulated process, they may rely on more abstract SynoPoints to do so, whereas when they need to perform more detailed analysis, they may opt for more detailed ones. However, switching instantly might incur a relatively high cognitive load, in order to mentally connect one SynoPoint to the next. Smooth animated transitions between representations may considerably ease this cognitive burden by visually connecting pairs of representations, allowing users to observe and keep track of the information conveyed by one representation when we switch to another~\cite{miao2019multiscale}, instead of having to make that connection purely mentally. As we are dealing with dynamic structures rather than static ones, however, we must also take time into account, which increases the complexity of the representations in general, and of the transitions in particular. Indeed, when switching from one SynoPoint to another, domain experts do not want to stop the playback, which would waste time and make the experience more tedious. So during the transition, we must keep communicating information about the DNA structure, including changes in position or bonding status, if at least one of two SynoPoints involved in the transition displays such properties. Here, we show that it is possible to design smooth transitions between the SynoPoints we implemented, allowing for a smooth and continuous traversal of the entire set of SynoPoints, from end to end.


\begin{figure*}
  \centering
  \includegraphics[width=1.0\linewidth]{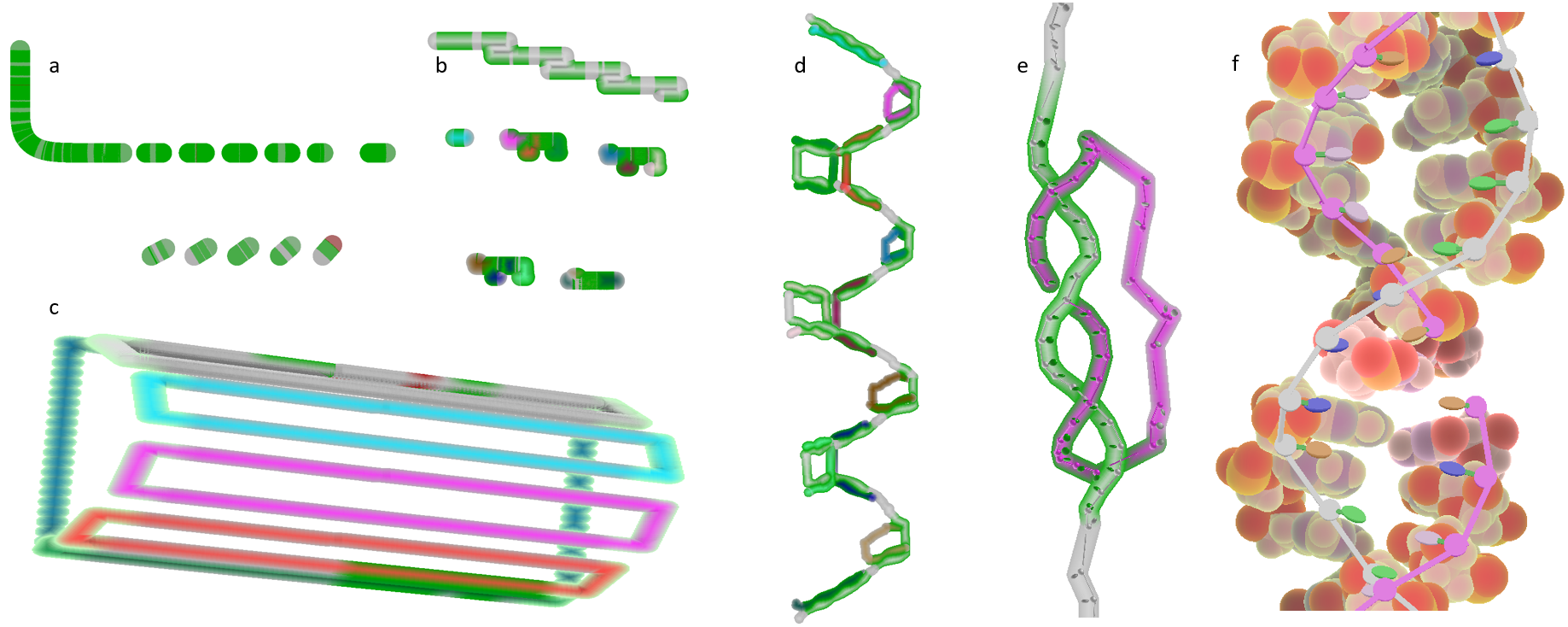}
  \caption{Animated transitions between consecutive levels in \textbf{SynopSet}. (a) Progress Bar $\triangleright$ Heatbar Array, the transition between~ autoref{fig:SynopSet}a and b; (b) Heatbar Array $\triangleright$ Schematic-2D, the transition between~\autoref{fig:SynopSet}b and c; (c) Schematic-2D $\triangleright$ Schematic-3D, the transition between~\autoref{fig:teaser}b3 and b4; (d) Schematic-3D $\triangleright$ Precise Snakes, the transition between~\autoref{fig:SynopSet}d and e; (e) Precise Snakes $\triangleright$ All-NT; f, All-NT $\triangleright$ Atomistic, the transition between~\autoref{fig:SynopSet}f and g} 
	\label{fig:SynopSetTransitions}
\end{figure*}

\textbf{Progress bar $\triangleright$ Heatbar Array} First, the Progress bar is split into several parts, one for each strand of DNA in the assembly. Each new strand ``crawls'' toward its new position, in a motion inspired by the Snake video game genre~\cite{demaria2002high}. Once the ``head'' of the strands have reached their destination, the rest of each strand is interpolated into place. Meanwhile, the coloring scheme gradually changes, through blending, from sequential coloring (where the bar fills with green from left to right and red from right to left) to heatbar coloring on the contour of each strand (where each nucleotide has its own color based on its bonding status); and to a unique identifier color for each strand. Thus, during the transition, both coloring schemes remain active and animated as the simulation progresses, until only one scheme is visible. Simultaneously, the shading smoothly transitions from flat to 3D. An intermediate state of such a transition is shown in~\autoref{fig:SynopSetTransitions}a, illustrating the transition between~\autoref{fig:SynopSet}a and b.
  
\textbf{Heatbar Array $\triangleright$ Schematic-2D} Each strand is bent into straight segments defined by their user-provided end points, with the segments being interpolated into their new positions. More specifically, for each control point on the spline that defines the representation of each strand, we consider a starting point, which is taken from the Heatbar Array, and an end point, taken from Schematic-2D; then, during the transition, an intermediate control point is linearly interpolated between the two. This process is applied to all control points, so that the entire spline is interpolated smoothly, as well as the Ballchain for the strand generated from it. The coloring scheme remains the same, so no transition is required for colors, though they remain animated during the transition, to indicate any changes in the bonding status of any of the nucleotides---see~\autoref{fig:SynopSetTransitions}b, showing the transition between~\autoref{fig:SynopSet}b and c.
  
\textbf{Schematic-2D $\triangleright$ Schematic-3D} This transition follows the same process as the previous one: the control points defining the splines for each strand are linearly interpolated, and therefore, so are the strands themselves. Visually, this smoothly transforms a 2D shape into a 3D one, where each strand may be further bent into more straight segments, depending on the 3D shape specified by the user. In this transition, there are no changes to the coloring scheme, and status colors on the contours keep being updated during it. An example is shown in~\autoref{fig:SynopSetTransitions}c, illustrating the transition between~\autoref{fig:teaser}b3 and b4.
  
\textbf{Schematic-3D $\triangleright$ Precise Snakes} This transition is done in two phases. In the first one, we compute the average position of all nucleotides. This position is used as the ``middle'' of the complete structure at this time (whether it is assembled or not). Then, we translate the Schematic-3D shape to this middle position, without altering its shape. Once it is in position, the second phase starts: we interpolate the Schematic-3D shape into the actual shape of each strand, as determined by the precise simulated position of each nucleotide. In practice, we have a starting set of control points corresponding to the Schematic-3D shape centered on the middle of the structure, and an ending set of control points corresponding to the (time-smoothed) position of each nucleotide. At every frame during this phase, we interpolate an intermediate set of control points between the starting and ending sets, and use it to generate Ballchains, creating the Snakes that make up the representation. Once more, the coloring scheme remains the same. During the transition, the ending set is continuously updated to reflect any changes in the positions of the strands. An example is shown in~\autoref{fig:SynopSetTransitions}d, illustrating the transition between~\autoref{fig:SynopSet}d and e.
  
\textbf{Precise Snakes $\triangleright$ All-NT} We first generate the ellipsoid-and-stick meshes inside the Snakes, and fully transparent, so that they are not visible. Initially, the backbone of the strand is generated with the nucleotides stuck to it instead of being at their normal distance, so that everything fits inside the Snakes. Then, gradually, we extend these nucleotides away from the backbone so that they seem to emerge out of the Snakes, while gradually fading them in, and fading out the Snakes, using transparency. This extension of the nucleotides away from the backbone is performed by gradually scaling the meshes that compose them and their link to the backbone, while adjusting their positions to maintain the classic double helix shape. Paired nucleotides can now be identified visually by their proximity to each other, but we still use color to make it easier to perceive (on the connections between backbone and nucleotide). The color transition is smooth, since the decreasing opacity of the Snakes and increasing opacity and width of the All-NT SynoPoint results in dynamic alpha blending. Since Precise Snakes feature temporal smoothing while the All-NT SynoPoint does not, in order to ensure that both representations are co-located during the transition, we in fact apply temporal smoothing to \emph{both} representations during the transition, while gradually decreasing the width of the sliding window down to zero, eventually removing temporal smoothing altogether. An example is shown in~\autoref{fig:SynopSetTransitions}e, illustrating the transition between~\autoref{fig:SynopSet}e and f.
  
\textbf{All-NT $\triangleright$ Atomistic} The transition from All-NT to Van der Waals (VdW) surfaces representing all atoms is done by fading in the VdW surfaces while fading out the All-NT structures (still with transparency). Since they are located in the same positions in time, with roughly similar shapes and sizes, this transition is smooth and aesthetically pleasing. The coloring scheme changes from nucleotide-specific colors to atom-specific colors (with halos for the bonding status), and the color transition is also handled through transparency. An example is shown in~\autoref{fig:SynopSetTransitions}f, illustrating the transition between~\autoref{fig:SynopSet}f and g.

\section{Animating the DNA Strands}
The Precise Snakes, All-NT and Atomistic SynoPoints all feature animation for the DNA strands, reproducing the motion from the simulation. However, due to the sheer amount of motion involved at even moderate playback speeds, merely displaying each simulated nucleotide where it is at any point in time would not be sufficient, as it would lead to very high-frequency motion, making it very difficult for domain experts to perceive anything clearly, and gain any insight from their analysis. At low playback speeds, on the other hand (below 30 simulation steps per second), motion would appear discrete, with objects moving instantaneously from one point to another at regular intervals, as opposed to continuous motion. Smoothing is therefore necessary for a comfortable experience, and time-efficient analysis thanks to high playback speeds.

\subsection{Smoothing within Periodic Boundary Conditions}
\label{sec:smoothing}
To generate continuous instead of discrete motion, when we compute a rendering frame between two simulation steps $S_{i}$ and $S_{i+1}$, where $S_{i}$ is at simulation time $T_{i}$ and $S_{i+1}$ is at simulation time $T_{i+1}$, we measure the current simulation time $t$, where $T_{i} \le t \le T_{i+1}$, to compute the proper linear interpolation weight. Then, for every nucleotide, we can linearly interpolate its position between the one it holds at step $S_{i}$ and the one it holds at $S_{i+1}$.

In addition, when rendering DNA using Precise Snakes, and in order to filter very high-frequency motion while preserving the general dynamics of the simulation, we use sliding window averaging as a form of temporal smoothing. At any time $t$, the position of a nucleotide is computed as the average of all the positions it takes between $t-R_{s}$ and $t+R_{s}$, where $R_{s}$ is the smoothing radius. Feedback from domain experts showed a smoothing radius of 12 simulation steps to be good compromise. It follows that with Precise Snakes, the interpolation done between steps $S_{i}$ and $S_{i+1}$ is in fact performed between smoothed positions $Smoothed_{i}$ and $Smoothed_{i+1}$, where $Smoothed_{i}$ features positions averaged over a sliding window ranging from $S_{i-12}$ to $S_{i+12}$, while $Smoothed_{i+1}$ contains positions averaged from $S_{i-11}$ to $S_{i+13}$.

Molecular dynamics simulation tools, including oxDNA, commonly use periodic boundary conditions (PBCs) for their simulation environment, in order to approximate a much larger system than what is actually simulated. It is essentially a finite volume where, if a particle exits from one side, it will reappear on the opposite side with the same velocity, just as in the Pacman video game. This presents a challenge for our Ballchain technique, because it relies on chains of impostor spheres interpolated between successive nucleotides within a given DNA strand. So if a strand features a nucleotide at position $(x,y,z) = (98,0,0)$ and another one at $(99,0,0)$, the space between these two positions will be filled with spheres. However, if their respective $x$ coordinates increase and the PBCs limit the value of $x$ to $100$, then the first nucleotide may find itself at $(99,0,0)$, while the second will end up at $(0,0,0)$. We would therefore have a Ballchain spanning the entire width of the PBCs in a straight line.

Further, if a given nucleotide holds positions with an $x$ value around $0$ or $100$ for a few successive simulation steps, temporal smoothing will put it around $x = 50$ instead, thereby completely misrepresenting the output of the simulation. Raw coordinates may be supplied directly by the simulation tool, or reconstituted in a pre-processing phase before visualization.

In order to solve these issues, we perform all smoothing operations on \emph{raw} coordinates instead of PBCs coordinates, where raw coordinates are defined as the positions that nucleotides would hold if no periodicity were applied. After smoothing these raw coordinates, which are of course devoid of any sudden jumps, we apply the PBCs constraints to put them within the proper limits. We further modify the Ballchain technique to ensure that if the distance between two successive nucleotides within a DNA strand is significantly greater than the normal distance (around $0.34nm$), then no balls are generated between them. This means that if a DNA strand crosses PBCs, it will appear in two pieces, one on each side, instead of spanning its entire width.

\subsection{Automated Camera Management}
While our more abstract SynoPoints feature no motion, Precise Snakes, the All-NT and Atomistic SynoPoints reproduce the motion from the MDS. Therefore, DNA strands can move around the scene, sometimes too fast for users to comfortably keep them in view through manual camera management. To mitigate this issue, we implemented automated camera management. While displaying moving DNA strands, we continuously compute the average position of all nucleotides, which we define as the center of the entire structure. Given this center, we then compute the radius of the smallest sphere that can encompass all DNA strands. Then, while maintaining the camera's orientation, so as to minimize confusion for the user, we move the camera so that it is facing the center of the structure, from a distance where the computed sphere fits precisely within the viewing frustum~\cite{kouvril2022molecumentary}. This continuous management keeps unexpected camera changes to a minimum, following the principle of least astonishment~\cite{saltzer2009principle}, while maintaining the entire structure in view at all times, and letting users rotate it at will. If they wish to control the camera manually, they can disable the automated camera management. 

\section{Results}
\label{Results}
We demonstrate the effectiveness of SynopSet on 12 simulations of DNA nanotechnology designs. In total, five nanostructures are investigated. A 2D plane composed of 768 nucleotides had been extensively studied by Snodin \emph{et al.}~\cite{snodin2016direct}. We name it \textbf{Snodin's Plane}. Even though it has only 768 nucleotides, it was and still is "the largest DNA nanostructure whose self-assembly has been directly simulated" and published. We have 4 trajectories generated for it: full assembly from individual strands, a detailed look at the last staple zipping event, disassembly by applying forces at both ends at 2 different temperatures. In the seminal paper that proposed oxDNA's force field~\cite{ouldridge2010dna}, a DNA nano tweezer composed of 84 NT had been studied by Ouldridge \emph{et al.} We name it \textbf{Ouldridge's Nanotweezer} and 2 trajectories have been generated for it: addition of the fuel strand to close the tweezer, and addition of the anti-fuel strand to open it. In a recent paper using DNA bricks, one of the many examples introduced by Wang \emph{et al.} is a tetrahedron composed of 396 NT. We name it \textbf{Wang's Tetrahedron} and 2 trajectories have been generated for it: full assembly from individual strands, and disassembly at high temperature ($ 100^\circ C $). A biomimetic nano channel composed of 300 NT had been designed and studied by Burns \emph{et al.}~\cite{burns2016biomimetic}. We name it \textbf{Burns' Pore} and 2 disassembly trajectories at different temperatures have been generated. Barišić \emph{et al.} modified the sequence of \textbf{Burns' Pore} to let it only have 5\% GC content, to study the effect of the proportion of CG content on the structure's stability. We name it \textbf{Barišić's Pore} and have generated for it 2 disassembly trajectories at different temperatures (the same as for \textbf{Burns' Pore}). All simulations have been carried out using oxDNA on clusters (Shaheen II and IBEX) provided by KAUST Supercomputing Lab, details of the simulations are mentioned below for each case. 


\subsection{Snodin's Plane: disassembly under force at  37 degrees C}

Snodin's Plane has a scaffold of 384 NT and 12 staples, each of 32 NT. For this run, both the first and the last NT of the scaffold are applied with two ``string'' \cite{Document25:online} forces along the Y axis: the first towards positive values, the second towards negative values. The temperature is at $ 37^\circ C $. In total, there are 10,000,000 ($10^{7}$) steps. The configuration (position of all NT) is recorded every $10^{3}$ steps. 

At the \textbf{Progress bar} level, one the bar gradually loses its green filling, meaning the hydrogen-bonds between base pairs are broken and the structure is disassembling. Soon after each drop of the green filling, the bar appears to refill with green but does not, meaning the hydrogen-bonds are then reformed but only temporarily. Red chunks occasionally appear and disappear at the right end, meaning misbonding also happens, but just temporarily. At frame 200 (which is equivalent to step 200 x $10^{3}$ = step $2.10^{5}$), about 10\% is disassembled (\autoref{fig:SynopSet}a). When we switch to the \textbf{Heatbar Array}, it shows that the broken bonds are sparsely dotted at similar distances on the scaffold strand, while as on the staple strands, the broken bonds are either at the ends or at the center (\autoref{fig:SynopSet}b). On the \textbf{Schematic-2D} level, a clear pattern can be immediately observed: the broken bonds are mostly at two thirds of each lane (\autoref{fig:SynopSet}c). As this structure is a plane which can be laid out just in 2D, the \textbf{Schematic-3D} level is almost the same as the \textbf{Schematic-2D} level. So we show a later frame, frame 2000 in~\autoref{fig:SynopSet}d. The disassembly progress is immediately seen and the ``two thirds pattern'' is even more prominent. With \textbf{Precise Snakes} (\autoref{fig:SynopSet}e), more detailed information is shown in an easily understandable way, especially if one has gone through the levels before. The ``two thirds pattern'' is caused by both ends of 4 staples attaching to the scaffold while their center is detached from the scaffold as indicated by the red arrows. The scaffold is almost in a straight line and the staples are hanging on the scaffold. Four staples are standing out, with their ends are attached to the scaffold while the center is left, forming a loop-like local structure. At the \textbf{All-NT} SynoPoint, after zooming into the region of interest based on prior findings at more abstract levels, the double helices are clearly shown. For example, 5 double helices can be seen in \autoref{fig:SynopSet}f. After further zooming in, the locations of the head and tail of the purple staple become clear. The NTs at the head, the tail, and the center are colored based on their types (A, T, C, G) because they are closer to the camera while the rest are colored based on their bonding states. A closer look at the head and tail at the \textbf{Atomistic} SynoPoint is shown in \autoref{fig:SynopSet}g. The occluded and non-occluded volume are clearly shown. The \textbf{SynopBond} summary figure (shown in the supplementary video), shows that even at the end of the simulation, all the staples are still attached to the scaffold. 

Figure 3 (the breaking of native base pairs tracking) from \cite{engel2018force} and figure 2 (kymograph of the assembly process) from Snodin \emph{et al.}'s work~\cite{snodin2016direct} are similar to the \textbf{SynopBond} summary figure. But they can only be viewed passively as static figures that are totally separate from the corresponding trajectories. By coupling the \textbf{Schematic-2D} SynoPoint with the \textbf{SynopBond} figure, way more information can be revealed at once. Take the above ``two thirds pattern'': it will only be manifested if each NT's pairing status is shown, so it will surely be averaged out from the strand or domain level bond statistics in the figures currently used in the domain. 
Figure 3C and 12 from \cite{engel2018force}, as well as Figure 4, 5, 9 from \cite{snodin2016direct} use the screenshots of all NTs' positions in the trajectory to show certain events. To digest the events, the viewer must go back and forth between the figures and text description and/or the summary statistics to make the connection. With the seamless transitions between each level in \textbf{SynopSet}, the requirements for such mental efforts to make the connection are not needed, or much lower. Besides, whole events can also be played at all different levels, which also facilitates the understanding of the dynamic behaviors. 

Figure 9 from \cite{snodin2016direct} also shows a nice abstraction to reveal the order of the assembly of all the staples. However, the resolution is at a sometimes higher than the domain level (the ``domain'' is defined in the paper and is half length of the staple). While in \textbf{SynopSet}, there is a whole hierarchy of abstraction levels, and all of them can be played at the resolution that can show each NT's bonding status. 

To summarize, in \textbf{SynopSet}, what can be achieved in the previous corresponding visualizations can be achieved better while the coupling of all the SynoPoints and SynopBond summary figure also provides the explanatory power that no previous tools have. 

\subsection{Other cases}

To reduce redundancy in text and figures, only the most prominent features that are not shown in the case above are listed here for all other cases' explanatory analysis, and the visuals for some of the cases are included in the supplementary video. For \textbf{Snodin's Plane: disassembly under force at 60 degrees Celius}, there are staples which totally fall off from the scaffold. After a strand is totally dropped from the scaffold, no refilling (re-forming of the hydrogen bond) is observed. This echoes what was categorized as ``local features'' in \cite{snodin2016direct} in a reverse manner. In the case of \textbf{Snodin's Plane: full assembly}, one interesting feature can be seen at this level: the bar is filled roughly chunk by chunk. Each chunk's beginning takes a while with the green bar's head shaking rapidly left and right, which identifies a hit and trial phase, and then suddenly the whole chunk gets green and never disassembled. Only this last sudden assembly is captured as ``local features'' in \cite{snodin2016direct} while the hit and trial phase is not mentioned. Notably, this full assembly data set is 37 GB, while the rendering and interaction are all in real-time. To our knowledge, no other tool can do this. \textbf{Snodin's Plane: the last staple zipping} is the final zipping step in the full assembly. It is singled out to allow users with hardware that has limited memory to try our tool. In the cases of \textbf{Ouldridge's Nanotweezer: closing} and {opening}, the conformation change of a nanostructure is demonstrated. Such changes are quite often the basis for the functional design of a structure. \textbf{Wang's Tetrahedron: full assembly} demonstrates the capability of showing a 3D structure made from DNA bricks. \textbf{Burns' Pore: disassembly at 85 degrees} and \textbf{at 100 degrees} demonstrate the disassembly of a 3D structure. One frame from $ 100^\circ C $ when one side of the pore is just disassembled is shown in~\autoref{fig:teaser}. The supplementary video has a detailed explanatory analysis demo for this structure. The same simulation settings for the \textbf{Burns' Pore} have also been performed for \textbf{Barišić's Pore}. All these 4 cases together showed that under the same sequence, the higher the temperature, the faster the disassembly is; and under the same topology but different sequence, the lower the GC content, the faster the disassembly is. All the cases here demonstrate \textbf{SynopSet}'s capability of handling a wide range different scenarios involving both DNA origami and DNA bricks structures.

\section{Domain Expert's Feedback}
Frequent interactions with domain experts were key to the design of SynopSet, providing invaluable data sets and guiding us through every decision. In this section, a domain expert who is among the authors of this paper offers his assessment of SynopSet.

In modern molecular biology, interdisciplinary research techniques and approaches have become  standard tools to investigate complex biological systems. MDS are an example where physicists, biologists, and computer scientists intensively collaborate to better understand molecular processes. With increasing computational performance possibilities, efforts are made to automate MDS workflows and to make them accessible to a wider pool of users and researchers not having the necessary hardware and software skills to setup such simulations ~\cite{poppleton2020design}. However, these new users are also not familiar with the classical data interpretation diagrams usually provided. In this work, we focused our research to develop novel, more intuitive data visualization concepts for coarse-grained MD simulations of DNA nanostructures. Especially, complex multilayer DNA nanostructures 
cannot be visually analyzed due to occlusion issues~\cite{ahmadi2020brownian}. Here, we overcome this problem using the multiscale model that allows the easy analysis and subsequent sequence optimizations of individual strands in the 2D view. The switch to the 3D view allows an estimation of the impact of non-optimal base-pairings on the global structure. In addition, the very promising computational approach to use oxDNA to visualize hybridization events of multiple oligonucleotides in nanostructures has the potential to strongly influence DNA nanotechnology research. An implementation of the presented technique in the standard design process will give researchers the possibility to quickly identify individual strands that cause miss-folding structures. In general, this work will set the foundations for novel data analysis tools for MD simulations because it enhances the understanding of the molecular interactions in complex structures due to intuitive visual concepts. It will also save researchers valuable time for the data analysis and facilitate the data interpretation for non-expert users.

\section{Formalization}

The representations in \textbf{SynopSet} rely on different idioms: Progress bars, Heatbars, Snakes, Ellipsoids-and-sticks, and Van der Waals surfaces. They also differ in granularity: the most prominent ones are the strand, nucleotide, and atom levels. Less obviously, they have different types of information layout as well: the Progress bar and Heatbar Array display information purely sequentially, with progress bars being sequential in nature, and the Heatbar Array simply displaying heatbars one after the other, in sequential order. In Schematic-2D and Schematic-3D, the layout is schematic, presenting simplified, idealized views of the structure. In Precise Snakes, All-NT and Atomistic, the layout is precise, as every element shown is displayed in its precise position, as per the simulation's output.

We observed that these properties (idiom, granularity, and information layout) could be thought of as separate axes, orthogonal to one another. As such, they generate a visualization space we are calling \textbf{SynopSpace}, owing to its ability to generate SynoPoints at various degrees of abstraction, suitable to the time-efficient communication of the \emph{story} of the simulation---its synopsis. In this space, each representation is a point, or \textbf{SynoPoint}. In the following subsections, the rationales behind each axis are laid out followed by the mapping between levels in the \textbf{SynopSet} and the \textbf{SynoPoints} as well as the proposed generalization of the space to other domains.

\begin{figure}[htp]
  \centering
  \includegraphics[width=0.7\linewidth]{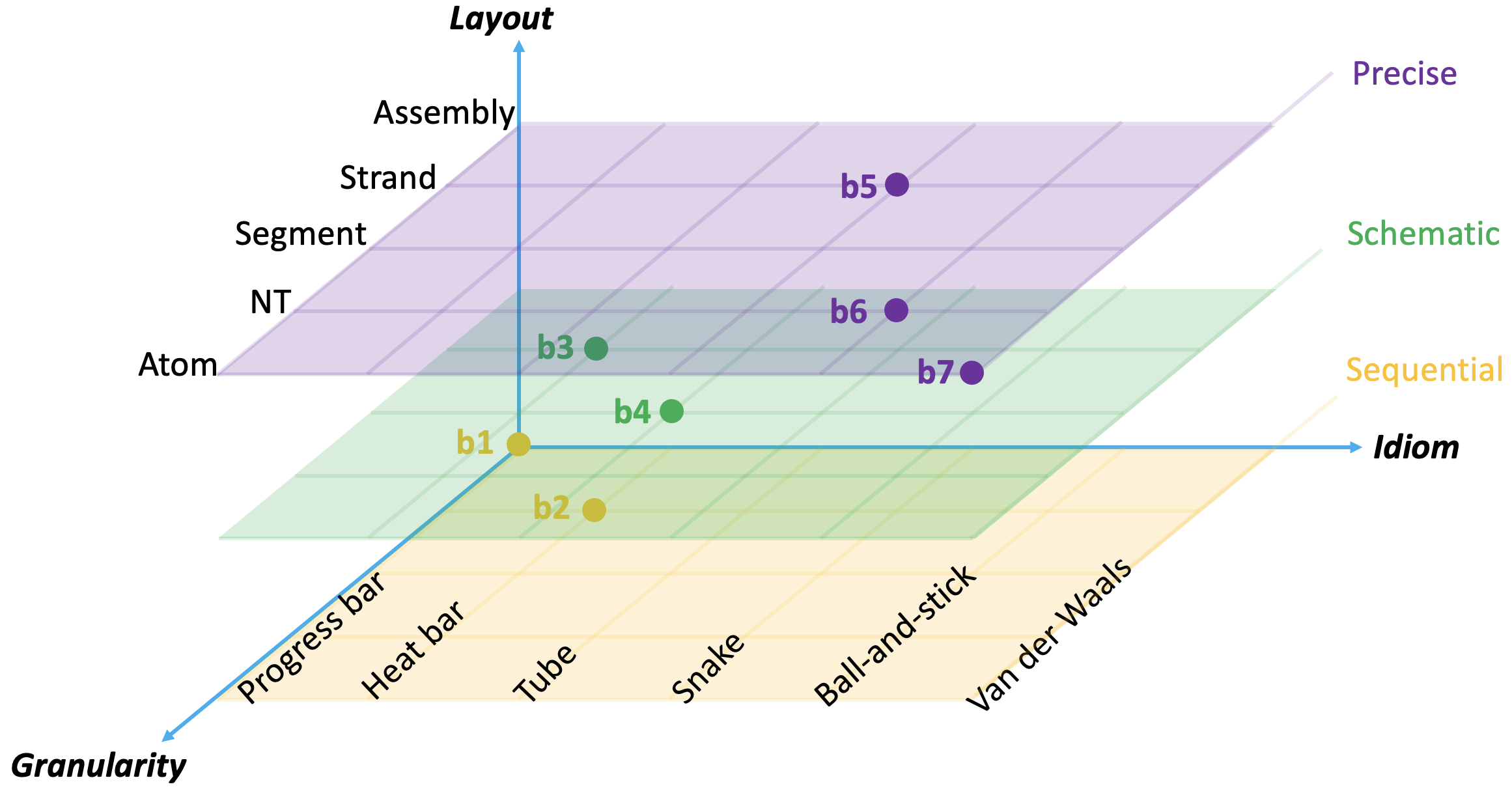}
  \caption{Construction of \textbf{SynopSpace} and the location of all \textbf{SynopSet} levels in the space. The space is constructed by using three axes: granularity, idiom, and layout. The three colored planes are three levels for the layout axis. Granularity levels are labeled along the edge of the purple plane. Idiom levels are labeled next to the yellow plane. In total, $5\times6\times3 = 90$ \textbf{SynoPoints} are in the space. The \textbf{Synopset} levels are highlighted by solid dots and labeled with b1-b7, corresponding to \autoref{fig:teaser}b1-b7 and \autoref{table:SynopSet_to_SynoPoints_mapping}}
	\label{fig:SynopSpace}
\end{figure}

\subsection{The Granularity Axis}

The \emph{Granularity} axis has most of its levels inherited from the domain. At the most abstract level stands one geometry that has all the information about an entire DNA nanostructure design aggregated into one holistic view. We call this level \textbf{Assembly}. One assembly is composed of individual strands. Those strands are usually the output from the deigning tool in the domain and also the ingredients needed to be synthesized to serve as the input for the wet lab (\emph{in vitro}). So, naturally, the next level is called \textbf{Strand}. In the DNA origami strategy, the long scaffold is folded by the short staples by forming crossovers. The crossover has one segment (one portion) of a strand paired with one region of the scaffold while the other segment of the same strand paired with another region of the scaffold. Similarly in the DNA bricks strategy, one brick can be viewed as several segments that serve as sticky ends to pair with other bricks. So, after the Strand level, we have \textbf{Segment}. Each strand, as well as each segment, is defined by its unique nucleotide sequence that is a permutation of four nucleotides A, T, C, G. We then have the nucleotide level, abbreviated as \textbf{NT} (as is common in genetics). The NT level is critical to domain experts because the DNA nanostructures suffer from a well-known problem: the designed structures’ yield is very low and one of the reasons is the non-intended base pairing that can be well-represented at the NT level. Each NT has a known chemical composition at the \textbf{Atomic} level. Sometimes, it is important to introduce modifications on a specific atom so as to make the entire assembly more stable or to give the assembly some specific function such as carrying a linker molecule at the 3' end, to have other molecules attached on the strand~\cite{burns2016biomimetic}. Further down the Atomic level, there are phenomena governed by quantum physics. As the computation demands associated with quantum physics simulations are huge, domain scientists rarely explore details down to such a level, and we have not addressed it. Similarly, on the other end, at the top of the Assembly level, there is a level similar to the quaternary structure of proteins that can describe the interaction between different DNA nanostructures. As this level goes beyond the the scope of an assembly process, we have not included it either. However, as domain practices advance on both ends of the \textit{Granularity} axis, it may be extended. 

\subsection{The Visual Idioms Axis}

To best convey the process at each granularity level, another axis that spans different \emph{Visual Idioms} is explored. To quickly grasp the current progress of the assembly simulation, the percentage that represents the intended and non-intended pairings can come in handy. We lean on the commonly used progress bar to present the overall progress of the simulation. We color it from one end in green to represent the percentage of the intended pairings, and from the other end in red to represent the percentage of the non-intended pairings. The other aspect in the space now shows up, the coloring scheme. Take the Progress bar: the just mentioned coloring method aggregates the emergent information at the Assembly granularity to both ends. If the viewer wants to see the progress for each strand, then the aggregation and the coloring can be performed on the Strand level, similarly for the Segment and the NT level. As demonstrated in the use cases, we are using the Progress bar at the Assembly granularity, i.e., one bar for the entire structure. We are just changing the coloring scheme to other granularities. Instead of a single bar, we can also use multiple bars (changing the geometry) when moving to other granularities. The coloring scheme can be viewed as another independent axis. However, the visualization space would then become very complicated and harder to navigate. We integrate the coloring scheme into the \emph{Visual Idiom} by adaptively choosing the most informative coloring scheme based on the current \emph{Visual Idiom} level and the \emph{Granularity} level. Now we move to the next idiom, the \textbf{Heatbar}. It encodes the information about each NT, whether it is intentionally paired to another NT or unintentionally paired, or not paired yet. While the progress bar and Heatbar can be depicted in 2D, in the end, we will need to depict each NT in its 3D position. Ensuring a smooth transition from 2D to 3D, we turn the 2D Heatbar idiom to the a 3D \textbf{Tube} idiom, then bend the straight Tube into the curved \textbf{Snake} idiom, which is used to represent the strand’s shape at its precise location. At the Tube idiom, the pairing status of each NT is still encoded by the color scheme. But at the Snake idiom, the pairing is also encoded by the vicinity between two strands. In other words, the Snake idiom allows for the perception of the typical double helix representation at the region where two strands are actually paired. Further down to the NT detail, we rely on the classical \textbf{Ball-and-stick} idiom. The phosphate sugar backbone of DNA is depicted in the beads (ball) on string (stick) manner. The base (nucleoside) is depicted by an ellipsoid close to the double helix middle line, which is then connected to the beads by another stick. This Ball-and-stick idiom encodes exactly the same amount of positional and orientational information that the oxDNA simulation model outputs. If there are non-intended pairings that would only be identifiable due to their red coloring in more abstract idioms, they can also be fundamentally revealed at this level by showing the actual base type. While the oxDNA model does not have the atomistic position information computed, it can be deduced from the NT position and orientation. The canonical \textbf{Van der Waals surface} idiom is used to depict the atoms. 

\subsection{The Layout Axis}

We have touched on the \emph{Layout} when introducing the Tube idiom, as the Tube can be viewed as the 3D version of the 2D Heatbar. More formally, we define the Progress bar and the Heatbar that has the pairing status of NT sequentially aligned from left to right as the \textbf{Sequential} layout. Sequential layout shows the overview information effectively and requires less time from the viewer to digest. When the time budget is bigger, more information such as the relationship between two strands in the target configuration can be shown, then we define the \textbf{Schematic} layout, in which each strand’s or segment’s starting and ending positions are close to or at their positions in the final design. The Schematic layout at the Segment Granularity for a nanostructure is designed to be the same as the widely used 2D DNA diagrams in the DNA nanotechnology domain, so that the domain experts can quickly capture the essence of the Schematic layout and the relationship between other layouts. The curved Snake idiom, Ball-and-stick idiom, and Van der Waals surface idiom are introduced at their corresponding precise positions, which we call \textbf{Precise} layout to allow users to examine the simulation in greater detail.

These three axes (~\autoref{fig:SynopSpace}a) will generate a space with 90 \textbf{SynoPoints} (\autoref{fig:SynopSpace}b). All levels from \textbf{SynopSet} are in this space (\autoref{fig:SynopSpace}c). By mapping each level in \textbf{SynopSet} to a \textbf{SynoPoint}, we introduce the formal names of each level. Take \textbf{Progress bar}: as the information is laid out sequentially, and it shows the overall progress of the assembly as a whole, it is formally \textbf{Sequential-Assembly-Progress bar}. Similarly, with \textbf{All-NT}, as the positions are all precise, each NT has a position, and the ball-and-stick idiom is used, it is formally \textbf{Precise-NT-Ball-and-stick}. Other mappings are given in \autoref{table:SynopSet_to_SynoPoints_mapping}. All 7 \textbf{SynoPoints} in \textbf{SynopSet} are highlighted in \autoref{fig:SynopSpace}c.

\begin{table}[]
\centering
\begin{tabular}{|l|l|l|}
\hline
\textbf{SynopSet Level} & \textbf{SynoPoint} & \textbf{Label*}                          \\ \hline
Progress bar   & Sequential-Assembly-Progress bar  & b1 \\ \hline
Heatbar Array  & Sequential-Strand-Heatbar & b2      \\ \hline
Schematic-2D   & Schematic-Strand-Heatbar  & b3      \\ \hline
Schematic-3D   & Schematic-Segment-Tube & b4            \\ \hline
Precise Snakes & Precise-Strand-Snake & b5              \\ \hline
All-NT         & Precise-NT-Ball-and-stick   & b6       \\ \hline
Atomistic      & Precise-Atom-Van der Waals surface & b7 \\ \hline
\end{tabular}
\caption{Map SynopSet levels to SynoPoints in SynopSpace} 
*Label is in \autoref{fig:SynopSpace}c. It also echoes to \autoref{fig:teaser}, b1-b7.
\label{table:SynopSet_to_SynoPoints_mapping}
\end{table}

\section{Conclusion}
DNA nanostructures are ever-increasing in complexity, and understanding their dynamic behavior is key to their research. In this work, we have proposed a visual abstraction set, \textbf{SynopSet}, that spans 7 representations (SynoPoints) to convey the dynamics of DNA nanotechnology simulations at multiple scales, to allow the interactive explanatory analysis of their simulations. The seamless transitions between those representations further helps the user to better understand the dynamic behavior of the MDS, with a lower cognitive burden, particularly to connect the information in the separate representations, thanks to smooth transitions. The most interesting events during a long simulation can thus be quickly identified and examined in further detail.

We formalized the description of the SynoPoints with a visualization space called \textbf{SynopSpace}. Its construction and design principles are introduced. We believe there are a variety of other processes that can make use of our concept of visualization space. Biology alone invites further application of SynopSpace as there are many different representations at different granularities, layouts and visual idioms. We submit that the space can be adapted to other domains by adapting the \emph{Granularity} axis to the domain-relevant concepts, and the \emph{Visual Idiom} to commonly used visual styles for those granularities. In protein simulation, for instnace, one would only need to change NT to Amino-Acid on the granularity axis to construct the space for it. For cell division as another example, the \textit{Granularity} axis could have: Cell, Compartment, Organelle, Macromolecule, Molecule, Atom; and the \emph{Visual Idiom} could be adapted from: Progress bar, Ellipsoid, Stick. 

Our approach for placing standard visualization techniques in the context of a larger space can lead to systematic discoveries of new visual representations. When inspecting the visualization space in detail, we might see that some idioms are strongly correlated with a certain layout or granularity. It might be that most of the idioms will remain bound by these correlations, but it might also be that some idioms escape their original ties and will form surprising idioms outside their original \emph{habitat}. For example, the progress bar itself was the highest abstraction, where no structural information is represented. However, here we morph our Progress Bar into multiple Heatbars, Tubes, and eventually Snakes, which is an entirely new way of using the concept of a progress bar. Without placing the progress bar into the three-dimensional visualization space, we might not have found its new promising extended form. Such research methodology applies to various other visualization scenarios where spanning a visualization space by recognizing important dimensions might lead to the development of new, surprisingly effective visual representations. While this work focuses on the visualization of the trajectories that are already generated through molecular dynamics simulations, the concept of \textbf{SynopSet} abstraction and \textbf{SynopSpace} can also be leveraged in the generation of animations for such processes, meaning that the complex behaviors can be first authored in more abstracted representations and then the more detailed animations will be automatically computed, so that the whole animation creation workflow can be made much more efficient. 

The large data sets\footnote{The tool and the data sets are available at https://github.com/nanovis/SynopSet.} presented in our work are another side-contribution to the visualization community. Such data sets can be used to test, for example, multiscale visualization, automated camera management systems, automated identification of events and smart labeling according to scales and events. 

\bibliographystyle{unsrt}
\bibliography{bibliography}

\end{document}